\documentstyle[12pt,epsf]{article}

\newcommand{\lsim}{\mathrel{\raisebox{-.6ex}{$\stackrel{\textstyle<}{\sim}$}}}

\def\beq{\begin{equation}}
\def\eeq{\end{equation}}
\def\bea{\begin{eqnarray}}
\def\eea{\end{eqnarray}}

\begin{document}

\thispagestyle{empty}

\font\fortssbx=cmssbx10 scaled \magstep2
\hbox to \hsize{
%\special{psfile=/NextLibrary/TeX/tex/inputs/uwlogo.ps
%                         hscale=8000 vscale=8000
%                         hoffset=-12 voffset=-2}
%\hskip.5in\raise.1in
\hbox{\fortssbx University of Wisconsin - Madison}
      \hfill$\vtop{
\hbox{\bf MADPH-99-1110}
\hbox{\bf AMES-HET-99-04}
\hbox{April 1999}}$ }

\vspace{.5in}

\begin{center}
{\large\bf Majorana neutrino masses from\\
neutrinoless double beta decay and cosmology}\\
\vskip 0.4cm
{V. Barger$^1$ and K. Whisnant$^2$}
\\[.1cm]
$^1${\it Department of Physics, University of Wisconsin, Madison, WI
53706, USA}\\
$^2${\it Department of Physics and Astronomy, Iowa State University,
Ames, IA 50011, USA}\\
\end{center}

\vspace{.5in}

\begin{abstract}				    

When three Majorana neutrinos describe the solar and atmospheric
neutrino data via oscillations, a nonzero measurement of neutrinoless
double beta ($0\nu\beta\beta$) decay can determine the sum of neutrino
masses $\sum m_\nu$ if the solar solution has small-angle mixing, and
place a lower bound on $\sum m_\nu$ for large-angle solar mixing. If in
addition a nonzero $\sum m_\nu$ is deduced from cosmology, the neutrino
mass spectrum may be uniquely specified for some ranges of neutrino
parameters. For $\sum m_\nu > 0.75$~eV, the small-angle solar solution
is excluded by the current upper limit on neutrinoless double beta
decay. In models with maximal solar mixing the $CP$ phases of the
neutrinos may be strongly constrained by stringent upper bounds on
$0\nu\beta\beta$ decay.

\end{abstract}

\thispagestyle{empty}
\newpage

\noindent
{\bf 1. Introduction.}  Recent results from
Super-Kamiokande~\cite{SuperKsol,SuperKatm} support neutrino
oscillation explanations of the solar~\cite{solarexp,solarth} and
atmospheric~\cite{atmosexp,atmosth,oldatmos} neutrino anomalies. Global
fits to all the data indicate that neutrino oscillations among three
neutrino species are sufficient to describe the solar and atmospheric
data~\cite{barhall,bw98}, and estimates have been obtained for the
neutrino mass-squared differences required to explain the data ($\delta
m^2_{atm} \sim 3\times 10^{-3}$~eV$^2$ for atmospheric
neutrinos~\cite{SuperKatm} and $\delta m^2_{sun} \sim
10^{-10}$~\cite{bw98,bks98,bw99} or $10^{-5}$~eV$^2$~\cite{bks98} for
solar neutrinos). However, neutrino oscillations only put restrictions
on the mass-squared differences, and do not constrain the absolute
neutrino mass scale; they also do not distinguish whether the smaller
mass splitting (the one responsible for solar neutrino oscillations) is
between the two largest mass eigenstates or the two smallest. In order
to learn about the actual neutrino masses, we must look elsewhere.

Studies of the power spectra of the cosmic microwave background
radiation and galaxies can provide information on the sum of the
neutrino masses, $\sum m_\nu$\cite{hdm,tegmark,lya}.  Another
possibility for learning about neutrino masses is neutrinoless double
beta ($0\nu\beta\beta$) decay~\cite{0nubb}, which can occur if massive
neutrinos are Majorana (as is nominally expected~\cite{books}), in which
case lepton number is not conserved. If three Majorana neutrinos are
nearly degenerate, a nonzero measurement of neutrino mass coupled with
an upper limit on $0\nu\beta\beta$ decay can place strong constraints on
the Majorana neutrino mixing matrix~\cite{majorana}. In this Letter, we
derive a simple formula that summarizes these constraints, and
generalize the argument to include the effects of mass splittings. We
find that a nonzero measurement of neutrinoless double beta
($0\nu\beta\beta$) decay would determine the sum of neutrino masses
$\sum m_\nu$ if the solar solution has small-angle mixing, and place a
lower bound on $\sum m_\nu$ for large-angle solar mixing. For $\sum
m_\nu > 0.75$~eV, the small-angle solar solution is excluded by the
current upper limit on neutrinoless double beta decay. Simultaneous
nonzero measurements of $0\nu\beta\beta$ decay and $\sum m_\nu$ may
uniquely specify the structure of the neutrino mass eigenstates for some
ranges of neutrino mass parameters. Furthermore, in models with maximal
solar mixing the $CP$ phases of the neutrinos may be strongly
constrained by stringent upper bounds on $0\nu\beta\beta$ decay.

\begin{figure}
\centering\leavevmode
\epsfxsize=4.5in\epsffile{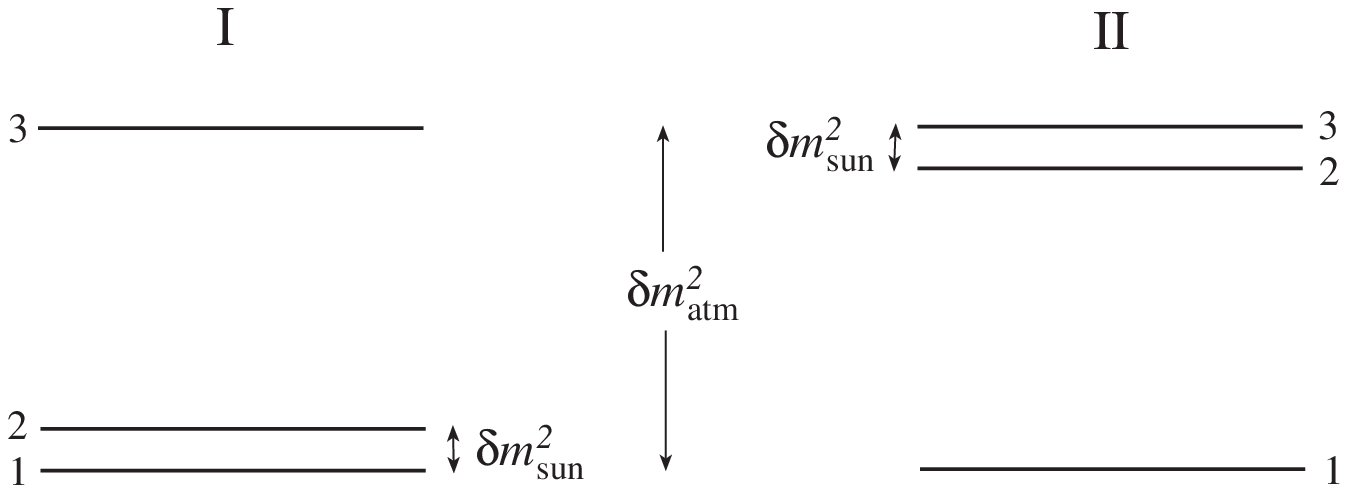}

\bigskip
\caption[]{\label{splitting} The two possibilites for the three-neutrino
mass spectrum.}
\end{figure}

\noindent
{\bf 2. Formalism.} We assume that there are only three active neutrino
flavors with Majorana masses, and that neutrino oscillations account for
the solar and atmospheric anomalies, with mass-squared differences
$\delta m^2_{sun} \ll \delta m^2_{atm}$ (our arguments hold for either
matter-enhanced~\cite{MSW} or vacuum solar~\cite{vacuum} solutions).
Assigning the mass eigenvalues $m_1 < m_2 < m_3$, there are two possible
mass spectra that can describe the oscillation data (see
Fig.~\ref{splitting})
\bea
\delta m^2_{21} &=  \delta m^2_{sun} \,, \qquad
\delta m^2_{32} =  \delta m^2_{atm} \,, \qquad {\rm~Spectrum~I} \,,
\label{model1}\\
\delta m^2_{21} &=  \delta m^2_{atm} \,, \qquad
\delta m^2_{32} =  \delta m^2_{sun} \,, \qquad {\rm~Spectrum~II} \,,
\label{model2}
\eea
where $\delta m^2_{jk} \equiv m_j^2 - m_k^2$. Spectrum~I (II) corresponds
to the case where the two closely degenerate states responsible for the
solar oscillation are the two smallest (largest) mass eigenstates.

\break
The charged-current eigenstates are related to the mass eigenstates by a
unitary transformation. In Spectrum~I we parametrize this transformation as
\beq
\left( \begin{array}{c} \nu_e \\ \nu_\mu \\ \nu_\tau \end{array} \right)
= U V \left( \begin{array}{c} \nu_1 \\ \nu_2 \\ \nu_3 \end{array} \right)
= \left( \begin{array}{ccc}
  c_1 c_3                           & c_1 s_3
& s_1 e^{-i\delta} \\
- c_2 s_3 - s_1 s_2 c_3 e^{i\delta} &   c_2 c_3 - s_1 s_2 s_3 e^{i\delta}
& c_1 s_2 \\
  s_2 s_3 - s_1 c_2 c_3 e^{i\delta} & - s_2 c_3 - s_1 c_2 s_3 e^{i\delta}
& c_1 c_2 \\
\end{array} \right) V
\left( \begin{array}{c}
\nu_1 \\ \nu_2 \\ \nu_3
\end{array} \right) \,,
\label{U}
\eeq
where $c_j \equiv \cos\theta_j$, $s_j \equiv \sin\theta_j$, and $V$ is
the diagonal matrix $(1,e^{i\phi_2},e^{i(\phi_3+\delta)}$). In
Eq.~(\ref{U}), $\phi_2$ and $\phi_3$ are additional phases for Majorana
neutrinos that are not measurable in neutrino oscillations; if $CP$ is
conserved, the phases in $UV$ are either $0$ or $\pi$. Then in
atmospheric and long-baseline experiments, the vacuum oscillation
probabilities are
\beq
P(\nu_\alpha \rightarrow \nu_\beta) =
4 |U_{\alpha3}|^2 |U_{\beta3}|^2 \sin^2 \Delta_{atm} \,, \qquad
\alpha \ne \beta \,,
\label{atmprob}
\eeq
where $\Delta_{atm} \equiv 1.27 (\delta m^2_{atm}/{\rm eV}^2) (L/{\rm
km})/(E/{\rm GeV})$ and terms involving the solar oscillation can be
ignored since they have not had time to develop. The solar $\nu_e$
vacuum oscillation probability is approximately given by
\beq
P(\nu_e \rightarrow \nu_e) = 1 - 2 s_1^2 c_1^2
- 4 c_1^4 s_3^2 c_3^2 \sin^2 \Delta_{sun} \,,
\label{sunprob}
\eeq
where $\Delta_{sun}$ is defined similarly to $\Delta_{atm}$ and the
oscillations involving $\Delta_{atm}$ have averaged: $\sin^2\Delta_{atm}
\rightarrow {1\over2}$. The CHOOZ reactor experiment~\cite{CHOOZ}
imposes the constraint
\beq
s_1 \lsim 0.23 {\rm~~~for~}\delta m^2_{atm} > 2\times10^{-3}{\rm~eV}^2 \,,
\label{chooz}
\eeq
but gives no limit for $\delta m^2_{atm} < 10^{-3}$~eV$^2$. The 504-day
atmospheric neutrino data imply~\cite{bw98}
\beq
s_1 < 0.3 {\rm~~~for~any~} \delta m^2_{atm} \,.
\label{s1limit}
\eeq
Thus in Spectrum~I there is very little mixing of $\nu_e$ with the heaviest
state; because of the small size of $s_1^2$, most of the solar $\nu_e$
depletion is due to the $\Delta_{sun}$ term for either matter-enhanced
or vacuum solar neutrino oscillations, and the fitted values for the
solar oscillation amplitude are not greatly affected by the particular
value of $s_1$~\cite{barhall,bw98}.

For Spectrum~II, the oscillation probabilities can be obtained simply by
interchanging the roles of $m_1$ and $m_3$. Then if $UV$
is obtained from Eq.~(\ref{U}) by interchange of the first and third
columns of $UV$, fits to oscillation data for Spectrum~II will give the
same values for the parameters $\theta_j$ and $\phi_k$ as those for
I. The limit on $s_1$ then implies that there is very little
mixing of $\nu_e$ with the lightest state in II.

In $0\nu\beta\beta$ decay, the decay rate depends on the
$\nu_e$--$\nu_e$ element of the neutrino mass matrix~\cite{0nubb}, which
is
\bea
M_{ee} &= c_1^2 c_3^2 m_1 + c_1^2 s_3^2 m_2 e^{i\phi_2}
+  s_1^2 m_3 e^{i\phi_3} \,, \qquad {\rm~(I)} \,,
\label{Mee1}\\
&= c_1^2 c_3^2 m_3 + c_1^2 s_3^2 m_2 e^{i\phi_2}
+  s_1^2 m_1 e^{i\phi_3} \,, \qquad {\rm~(II)} \,.
\label{Mee2}
\eea
The form of $M_{ee}$ necessarily implies
\beq
|M_{ee}| \le m_3 \,,
\label{Meemax}
\eeq
where $m_3$ is the largest neutrino mass eigenvalue.
The recently improved 90\%~C.L. upper bound  on $M_{ee}$ from
$0\nu\beta\beta$ decay experiments is~\cite{Mlimit}
\beq
|M_{ee}| < 0.2 {\rm~eV} \,.
\label{Mlimit}
\eeq
The GENIUS experiment is anticipated to be sensitive to $|M_{ee}|$ as
low as 0.01~eV~\cite{GENIUS}.

Finally, since individual masses are not in general directly measurable,
more appropriate variables are the sum of the neutrino masses, $\sum
m_\nu$, and neutrino mass-squared differences. For three nearly
degenerate neutrinos the sum of neutrino masses is approximately given
by
\beq
\sum m_\nu \simeq 3 m_1 \,.
\label{sum0}
\eeq
A measurement of the cosmological power spectrum from (i) the cosmic
microwave background radiation by MAP~\cite{MAP} and
PLANCK~\cite{PLANCK}, (ii) red-shift surveys by the Sloan Digital Sky
Survey (SDSS)~\cite{SDSS} and the Two-Degree Field (2dF)~\cite{2dF},
and (iii) the Ly$\alpha$ forest of neutral hydrogen absorption in quasar
spectra~\cite{lya} may be sensitive to $\sum m_\nu$ as low as
0.4~eV~\cite{tegmark}.

\noindent
{\bf 3. Limits on the solar oscillation amplitude when \boldmath{$\sum m_\nu >{}$}0.45~eV.} The neutrino mass eigenstates will be nearly degenerate if the
mass eigenvalues
\beq
m_1 \simeq m_2 \simeq m_3 \,,
\label{m}
\eeq
are large with respect to the mass splittings $\delta m^2_{jk}$. In this
event, the mass splitting between the smallest and largest masses is
\beq
m_3 - m_1 = {\delta m^2_{31} \over (m_3 + m_1)}
\simeq {\delta m^2_{atm} \over 2m_1}
\simeq {0.0035{\rm~eV}^2\over 2m_1} \,.
\label{split}
\eeq
This mass difference is at least an order of magnitude smaller than
$m_1$ for $m_1 > 0.15$~eV, i.e., for $\sum m_\nu > 0.45$~eV. For such
small mass splittings the $0\nu\beta\beta$ decay limit in both Spectra~I
and II can be written
\beq
\left| c_1^2 c_3^2 + c_1^2 s_3^2 e^{i\phi_2} +  s_1^2 e^{i\phi_3} \right|
= {|M_{ee}|\over m_1} \le {|M_{ee}|_{max}\over m_1} \,.
\label{limit}
\eeq

\begin{figure}
\centering\leavevmode
\epsfxsize=4.5in\epsffile{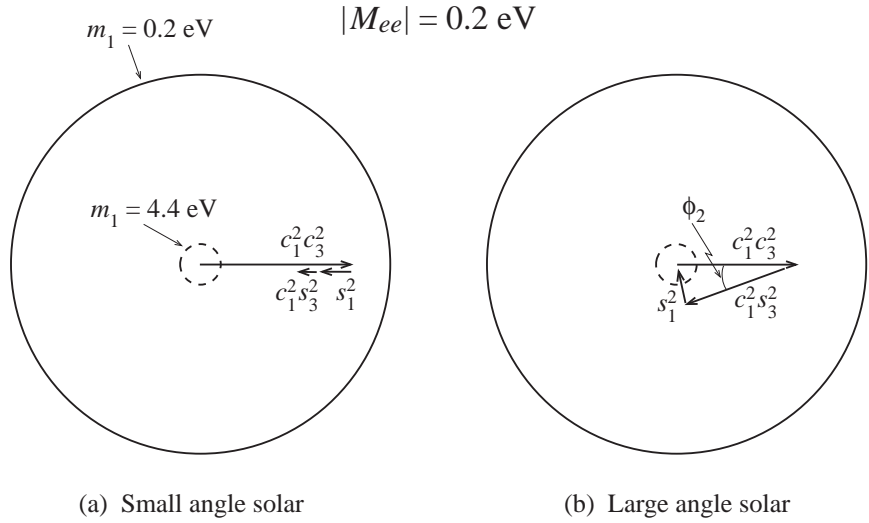}

\bigskip
\caption[]{\label{diagram} Typical examples of the $|M_{ee}|/m_1$ bound
(with $|M_{ee}|<0.2$~eV) for three nearly degenerate Majorana neutrinos
for the (a) small-angle and (b) maximal-mixing solar solutions. The
circles are the bounds assuming $m_1 = 0.2$~eV (solid) and $4.4$~eV
(dashed), where $m_1 = \sum m_\nu/3$. In (a) the bound is satisfied for
$m_1=0.2$~eV but not for $m_1=4.4$~eV; in (b) the bound is satisfied in
both $m_1$ cases for $\phi_2\approx \pi$.}
\end{figure}

The left-hand side of Eq.~(\ref{limit}) may be represented by the sum
of three complex vectors whose directions (the phase angles) are unknown
but whose lengths are determined by the mixing matrix parameters. A
geometric interpretation of the constraint in Eq.~(\ref{limit}) is that
the longest side minus the sum of the two shorter sides must be less
than $|M_{ee}|_{max}/m_1$. Given the current limits on $s_1$, one of the
two sides $c_1^2c_3^2$ or $c_1^2s_3^2$ must be the longest; without loss
of generality, we assume $c_3 > s_3$, so that $c_1^2 c_3^2$ is the
longest side. This limit is represented diagrammatically in
Fig.~\ref{diagram} for both the small-angle and maximal-mixing solar
solutions for two different values of $m_1$: $m_1=0.2$~eV (the current
upper limit on $|M_{ee}|$) and $m_1=4.4$~eV (the upper limit on $m_1$
from tritium beta decay measurements~\cite{tritium,inferred}).

Algebraically the constraint of Eq.~(\ref{limit}) may be written
\beq
c_1^2c_3^2 - c_1^2s_3^2 - s_1^2 \le {|M_{ee}|_{max}\over m_1} \,.
\label{limit2}
\eeq
Then using Eq.~(\ref{sum0}), the $0\nu\beta\beta$ decay limit becomes
\beq
s_3^2 \ge {1 - 2s_1^2 - {3|M_{ee}|_{max}\over \sum m_\nu}\over2c_1^2} \,.
\eeq
This in turn implies that the solar $\nu_e \rightarrow \nu_e$
oscillation amplitude is constrained by
\beq
A^{ee}_{sun} \equiv 4 c_1^4 s_3^2 c_3^2 \ge
1 - \left({3|M_{ee}|_{max}\over \sum m_\nu}\right)^2
- 2 s_1^2 \left(1 + {3|M_{ee}|_{max}\over \sum m_\nu} \right) \,. 
\label{Alimit}
\eeq
The same result is obtained with $s_3 > c_3$.

\begin{figure}
\centering\leavevmode
\epsfysize=5in\epsffile{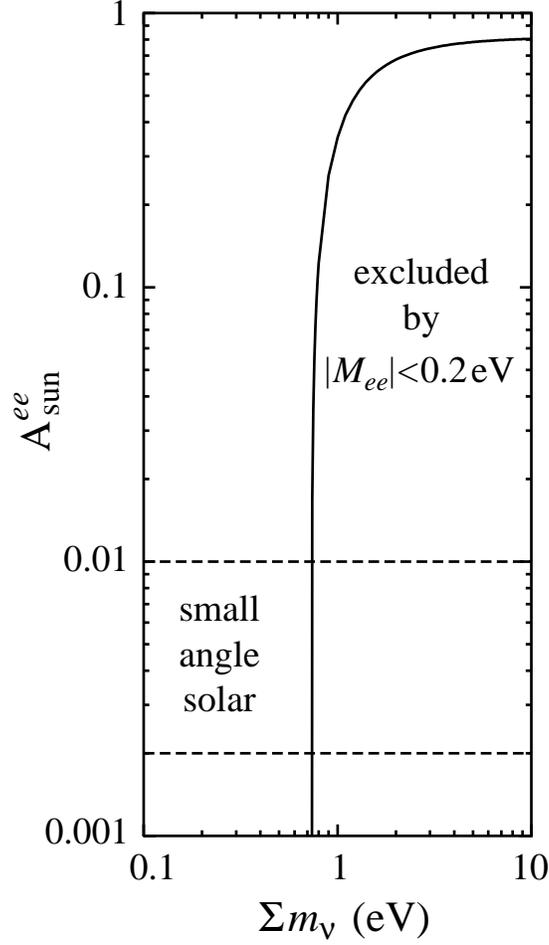}

\bigskip
\caption[]{\label{sunamp} Lower bound on the solar oscillation amplitude
$A^{ee}_{sun}$ versus the sum of the neutrino masses $\sum m_\nu$ based
on the current bound from $0\nu\beta\beta$ decay, $|M_{ee}|<0.2$~eV.
Spectra~I and II give nearly identical results. The allowed
range~\cite{bks98} for $A^{ee}_{sun}$ for the small-angle
matter-enhanced solar solution is also shown.}
\end{figure}

For any value of $\sum m_\nu > 3 |M_{ee}|_{max}/(1-2s_1^2)$ there will
be a lower limit on the size of $A^{ee}_{sun}$ from Eq.~(\ref{Alimit});
the most conservative limit occurs for the maximum value of $s_1^2$. In
Fig.~\ref{sunamp} we plot the lower limit on $A^{ee}_{sun}$ versus $\sum
m_\nu$ for $|M_{ee}|_{max} = 0.2$~eV and the current upper bound of $s_1
= 0.3$. For large $\sum m_\nu$ the lower bound on $A^{ee}_{sun}$
approaches $1 - 2 (s_1^2)_{max} \simeq 0.82$.

Given the current upper bound on $|M_{ee}|$ from $0\nu\beta\beta$ decay,
Fig.~\ref{sunamp} shows that for $\sum m_\nu \ge 0.75$~eV, the
small-angle matter-enhanced solar solution is excluded for three nearly
degenerate Majorana neutrinos. If the upper bound on $s_1$ were to
become more stringent, the limit in Eq.~(\ref{Alimit}) would be
tightened, and the small solar mixing solution would be excluded for
$\sum m_\nu$ smaller than 0.75~eV. Vacuum solar solutions, which have
$A^{ee}_{sun} = 0.6$--$1.0$~\cite{bw98,bks98,bw99}, are allowed.

The $0\nu\beta\beta$ constraint may be understood qualitatively as
follows. If there is small mixing of $\nu_e$ with two of the mass
eigenstates, then there is one dominant $U_{ej}$, in which case it is
impossible to have the three contributions to $M_{ee}$ combine to give a
small result for $|M_{ee}|$ if the individual neutrino masses are
greater than $|M_{ee}|$; see Fig.~\ref{diagram}a. On the other hand,
with large-angle solar mixing $\nu_e$ is a roughly equal mixture of two
eigenstates, and the three contributions to $M_{ee}$ can give a much
smaller result. In fact, $M_{ee}=0$ is always possible if $s_3=c_3$
(maximal solar mixing); see Fig.~\ref{diagram}b. Although a stringent
upper bound on $|M_{ee}|$ does not rule out Majorana neutrinos when
$s_3=c_3$, it does put a very tight limit on the Majorana phase angle
$\phi_2$. This is illustrated in Fig.~\ref{diagram}b, where a very small
$|M_{ee}|$ and small $s_1^2$ imply that the Majorana phase angle
$\phi_2$ is close to $\pi$. Such a value for $\phi_2$ is a natural
consequence of $CP$ conservation if $\nu_1$ and $\nu_2$ ($\nu_2$ and
$\nu_3$) have opposite $CP$ eigenvalues in Spectrum~I (II)~\cite{phases}.

\noindent
{\bf 4. Constraints on the neutrino mass spectrum for arbitrary \boldmath{$\sum
m_\nu$}.} For $\sum m_\nu < 1$~eV, the small splitting of the neutrino
masses indicated by the atmospheric and solar experiments can affect the
limit in Eq.~(\ref{Alimit}).  Writing Eqs.~(\ref{Mee1}) and (\ref{Mee2})
in terms of $m_1$ and $\delta m^2_{atm}$, in the limit that $\delta
m^2_{sun}$ can be ignored we find
\bea
M_{ee} &=& c_1^2 c_3^2 m_1
+ c_1^2 s_3^2 m_1 e^{i\phi_2}
+ s_1^2 \sqrt{m_1^2 + \delta m^2_{atm}} e^{i\phi_3} \,,
\label{Mee11}\\
\sum m_\nu &=& 2 m_1 + \sqrt{m_1^2 + \delta m^2_{atm}} \,,
\label{sum1}
\eea
in Spectrum~I, and
\bea
M_{ee} &=& c_1^2 c_3^2 \sqrt{m_1^2 + \delta m^2_{atm}}
+ c_1^2 s_3^2 \sqrt{m_1^2 + \delta m^2_{atm}} e^{i\phi_2}
+  s_1^2 m_1 e^{i\phi_3} \,,
\label{Mee22}\\
\sum m_\nu &=& m_1 + 2 \sqrt{m_1^2 + \delta m^2_{atm}} \,,
\label{sum2}
\eea
for Spectrum~II. Furthermore, the structure of the mass spectrum
requires
\bea
\sum m_\nu &>& \sqrt{\delta m^2_{atm}} \,, {\rm~~(I)} \,,
\label{summin1}\\
\sum m_\nu &>& 2 \sqrt{\delta m^2_{atm}} \,, {\rm~~(II)} \,.
\label{summin2}
\eea

{\it (a) Small-angle solar mixing.} For the small-angle
solar solution (which has $2\times10^{-3}
\le A^{ee}_{sun} \le 10^{-2}$~\cite{bks98} and thus $0.02 \le s_3
\le 0.05$), $s_3^2$ is negligible in Eqs.~(\ref{Mee11}) and
(\ref{Mee22}); then the allowed ranges for $|M_{ee}|$ are
\bea
\left| c_1^2 m_1 - s_1^2 \sqrt{m_1^2+\delta m^2_{atm}} \right| \  \le
&|M_{ee}|& \le \ 
 c_1^2 m_1 + s_1^2 \sqrt{m_1^2+\delta m^2_{atm}} \,, {\rm~(I)} \,,
\label{Mrange1}\\
c_1^2 \sqrt{m_1^2+\delta m^2_{atm}} - s_1^2 m_1 \ \le &|M_{ee}|& \le \ 
c_1^2 \sqrt{m_1^2+\delta m^2_{atm}} + s_1^2 m_1 \,, {\rm~(II)} \,.
\label{Mrange2}
\eea

\begin{figure}
\centering\leavevmode
\epsfysize=5in\epsffile{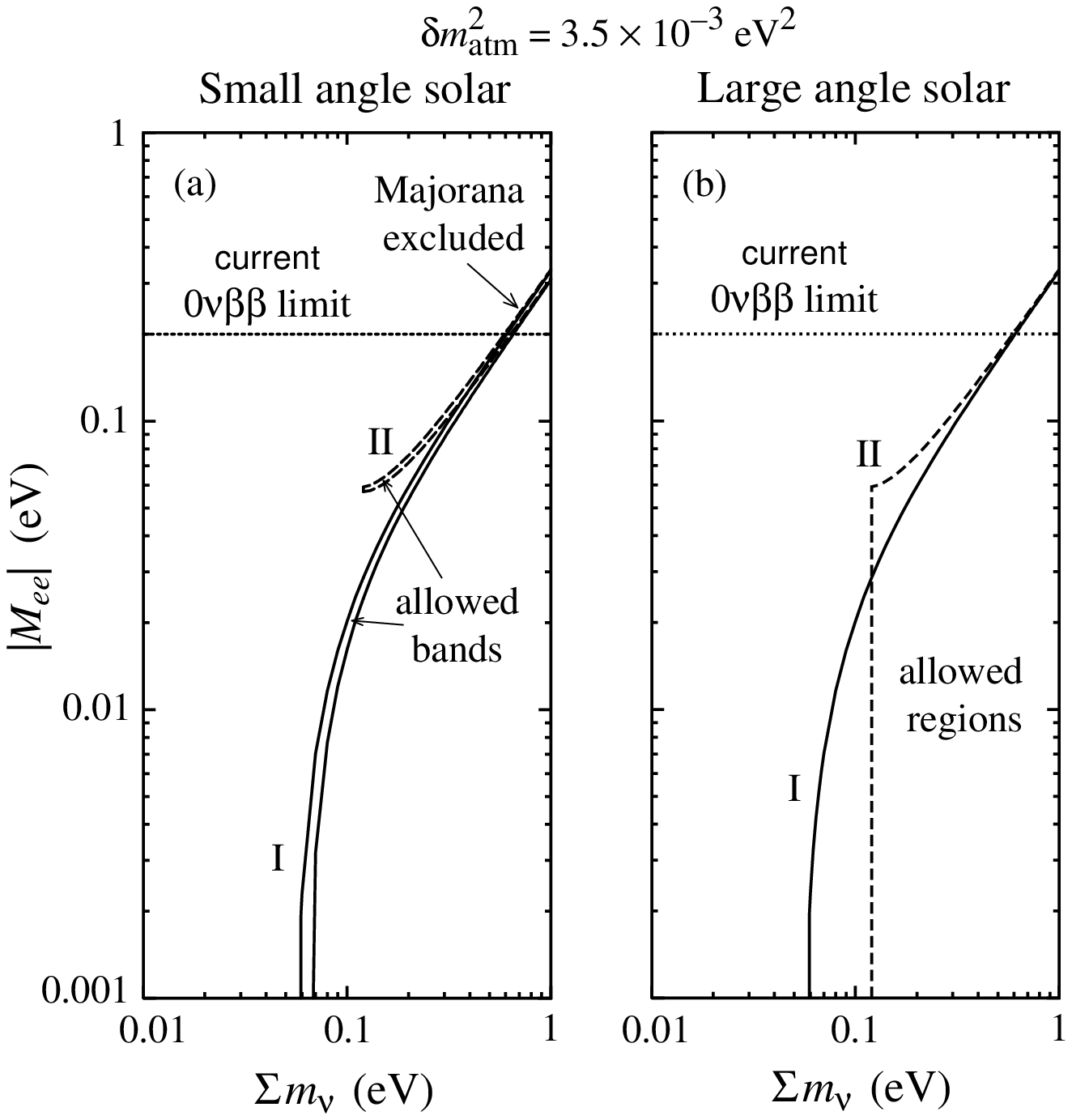}

\bigskip
\caption[]{\label{Mrange} Allowed regions of $|M_{ee}|$ versus the sum of
the neutrino masses $\sum{m_\nu}$ with $\delta m^2_{atm} =
3.5\times10^{-3}$~eV$^2$ for the (a) small-angle  and (b) large-angle
solar solutions. The mixing parameter $s_1$ has been varied over the
values $0 \le \sin\theta_1 \le 0.18$ allowed by the CHOOZ
constraint~\cite{CHOOZ}. Results are shown for Spectra~I (solid curves)
and II (dashed).}
\end{figure}

The allowed bands for $|M_{ee}|$ are shown in Fig.~\ref{Mrange}a
versus $\sum m_\nu$ (which is related to $m_1$ via Eqs.~(\ref{sum1}) and
(\ref{sum2})) in Spectra~I and II for $\delta m^2_{atm} =
3.5\times10^{-2}$~eV$^2$ and $0 \le s_1 \le 0.18$ (the CHOOZ constraint
for that $\delta m^2_{atm}$). Since both $s_1^2$ and $s_3^2$ are small
in the small-angle solar solution, only one mass eigenstate contributes
significantly to $M_{ee}$ and there is nearly a one-to-one
correspondence between $|M_{ee}|$ and $m_1$ (and hence between
$|M_{ee}|$ and $\sum m_\nu$):
\bea
|M_{ee}| &\simeq m_1 \,, \qquad {\rm(I)} \,,
\label{corresp1}\\
|M_{ee}| &\simeq \sqrt{m_1^2 + \delta m^2_{atm}} \,, \qquad {\rm(II)} \,.
\label{corresp2}
\eea
This relation between $|M_{ee}|$ and $m_1$ for the
small-angle solar solution implies that if a nonzero $|M_{ee}|$ is
measured, all of the neutrino masses will be determined for either
mass spectra in Fig.~\ref{splitting}.

The current limit $|M_{ee}| < 0.2$~eV in Fig.~\ref{Mrange}a shows that
for $\sum m_\nu > 0.75$~eV the small-angle solar solution is ruled out
for Majorana neutrinos in both Spectra~I and II. Although the range of
allowed $|M_{ee}|$ versus $\sum m_\nu$ expands if the presently allowed
range of $\delta m^2_{atm}$ from the Super-K atmospheric data is used,
the qualitative behavior of the allowed regions for Spectra~I and II
remains the same. Improved limits on $s_1$ and $\delta m^2_{atm}$ will
shrink the allowed ranges for the small-angle solar solution. The value
of $\delta m^2_{atm}$ can be more precisely measured in the
K2K~\cite{K2K} and MINOS~\cite{MINOS} long-baseline experiments.

It is evident from Fig.~\ref{Mrange}a that future measurements of
$|M_{ee}|$ and $\sum m_\nu$ could rule out the small-angle solar
solution for Majorana neutrinos in one or both of the mass spectra
possibilities. For example, Spectrum~II is ruled out for any $\sum
m_\nu$ when $|M_{ee}| < 0.05$~eV, and both spectra are excluded if,
e.g., $\sum m_\nu > 0.4$~eV and $|M_{ee}| < 0.1$~eV.  Alternatively,
nonzero measurements for both $|M_{ee}|$ and $\sum m_\nu$ could
distinguish between the two mass spectra.

{\it (b) Large-angle solar solution.} For the large-angle vacuum or
matter-enhanced solar solutions, the allowed range of $|M_{ee}|$
expands considerably. For the vacuum solar solution, the solar
oscillation amplitude is large, and may be maximal. Vacuum solutions
that allow maximal mixing can never be ruled out simply by lowering the
limit on $|M_{ee}|$. However, it may still be possible to distinguish
Spectrum~I from II.

The largest possible $|M_{ee}|$ occurs when all terms in
Eqs.~(\ref{Mee11}) and (\ref{Mee22}) add in phase, and is given by the
upper limits in Eqs.~(\ref{Mrange1}) and (\ref{Mrange2}). Then given
$\delta m^2_{atm}$ and Eqs.~(\ref{sum1}) and (\ref{sum2}), the largest
possible $|M_{ee}|$ can be found for a given $\sum m_\nu$; these results
are shown in Fig.~\ref{Mrange}b for $\delta m^2_{atm} =
3.5\times10^{-3}$~eV$^2$. Figure~\ref{Mrange}b shows that a nonzero
measurement on $|M_{ee}|$ implies a lower limit on $\sum m_\nu$; for
example, $|M_{ee}|=0.06$~eV implies that $\sum m_\nu > 0.20 (0.12)$~eV
in Spectrum~I~(II).

As can be seen in Fig.~\ref{Mrange}b, there are certain values of
$|M_{ee}|$ and $\sum m_\nu$ that are possible only for Spectrum~I, and
others that are possible only for II. In either of these cases a unique
mass spectrum could be selected; this conclusion follows from the fact
that $\sum m_\nu$ must always be greater than 2$\sqrt{\delta m^2_{atm}}$
in Spectrum~II, whereas in I it can be less.  Furthermore, since there
are two larger masses in II and only one larger mass in I, $|M_{ee}|$
can be larger in II than in I for the same value of $\delta
m^2_{atm}$. There is also a large region in Fig.~\ref{Mrange}b that
could be obtained in either I or II, in which case the mass spectra
would not be differentiated by the $0\nu\beta\beta$ and $\sum m_\nu$
measurements. The allowed regions for $|M_{ee}|$ expand as $\delta
m^2_{atm}$ is varied over its presently allowed range, but there is
still considerable area unique to each spectrum.

The allowed regions for large solar mixing are obtained without using
any information about $\theta_3$. If $s_3 = c_3$ (which corresponds to
maximal solar mixing), then as noted above the lower bound on $|M_{ee}|$
is zero. A precise determination of $s_3$ could reduce the allowed
ranges of $|M_{ee}|$ versus $\sum m_\nu$.

\noindent
{\bf 5. Summary.} Our main conclusions regarding models with three
Majorana neutrinos are as follows:

(i) For the small-angle matter-enhanced solar solutions there is an
approximate relation between $|M_{ee}|$ and $\sum m_\nu$, which implies
that a nonzero measurement of $|M_{ee}|$ determines $\sum m_\nu$ in
these models. If the sum of neutrino masses is determined by
cosmological power spectra measurements to be greater than about
0.75~eV, then the small-angle solar solution is ruled out by the
current stringent limit on neutrinoless double beta decay. Further
improvement of the $0\nu\beta\beta$ decay limit could rule out the small-angle
solar solution at even smaller nonzero values of $\sum m_\nu$, or
perhaps distinguish between Spectrum~I (in which the two lightest mass
eigenstates are responsible for the solar oscillation) and Spectrum~II
(the two heaviest mass eigenstates are responsible for the solar
oscillation).

(ii) Large-angle vacuum oscillation scenarios are largely safe from
$0\nu\beta\beta$ decay experimental constraints. However, a nonzero
measurement of $|M_{ee}|$ places a lower limit on $\sum m_\nu$ in these
models. The next generation of experiments measuring $|M_{ee}|$ and
$\sum m_\nu$ may be able to distinguish between Spectra~I and II, and
could perhaps give information on relative $CP$ phases of the neutrino
mass eigenstates. In particular, in models with maximal solar mixing,
$|M_{ee}|$ substantially below $\sum m_\nu/3$ is only possible when
the two neutrino mass eigenstates primarily contributing to $M_{ee}$
have $CP$ phases that differ by about $\pi$.

\medskip\noindent 
{\bf Acknowledgements.}  This work was supported in part by the
U.S. Department of Energy, Division of High Energy Physics, under Grants
No.~DE-FG02-94ER40817 and No.~DE-FG02-95ER40896, and in part by the
University of Wisconsin Research Committee with funds granted by the
Wisconsin Alumni Research Foundation.

\clearpage

% Fig. 1

% Fig. 2

% Fig. 3

% Fig. 4

\newpage

\begin{thebibliography}{99}
%
\bibitem{SuperKsol}
Super-Kamiokande Collaboration,
Y. Fukuda {\it et al.}, Phys. Rev. Lett. {\bf 82}, 1810 (1999);
Phys. Rev. Lett. {\bf 82}, 2430 (1999);
M.B. Smy, hep-ex/9903034,
talk at {\it DPF-99}, Los Angeles, California, January 1999;
G. Sullivan, talk at Aspen Winter Conference, January, 1999.
%
\bibitem{SuperKatm}
Super-Kamiokande Collaboration,
Y. Fukuda {\it et al.}, Phys. Rev. Lett. {\bf 81}, 1562 (1998);
Phys. Rev. Lett. {\bf 82}, 2644 (1999).
%
\bibitem{solarexp}
B.T. Cleveland {\it et al.}, Nucl. Phys. B (Proc. Suppl.) {\bf 38}, 47
(1995);
GALLEX collaboration, W. Hampel {\it et al.}, Phys. Lett. {\bf B 388},
384 (1996);
SAGE collaboration, J.N. Abdurashitov {\it et al.}, Phys. Rev. Lett.
{\bf 77}, 4708 (1996);
Kamiokande collaboration, Y. Fukuda {\it et al.}, Phys. Rev. Lett,
{\bf 77}, 1683 (1996);
%
\bibitem{solarth}
See, e.g., J.N. Bahcall, S. Basu, and M.H. Pinsonneault, Phys. Lett.
{\bf B 433}, 1 (1998), and references therein.
%
\bibitem{atmosexp}
Kamiokande collaboration, K.S. Hirata {\it et al.}, Phys. Lett.
{\bf B 280}, 146 (1992); Y. Fukuda {\it et al.}, Phys. Lett.
{\bf B 335}, 237 (1994);
IMB collaboration, R. Becker-Szendy {\it et al.}, Nucl. Phys. Proc.
Suppl. {\bf 38 B}, 331 (1995);
Soudan-2 collaboration, W.W.M. Allison {\it et al.}, Phys. Lett. {\bf
B 391}, 491 (1997);
MACRO collaboration, M. Ambrosio {\it et al.},
Phys. Lett. {\bf B 434}, 451 (1998).
%
\bibitem{atmosth}
G. Barr, T.K. Gaisser, and T. Stanev, Phys. Rev. {\bf D 39},
3532 (1989);
M. Honda, T. Kajita, K. Kasahara, and S. Midorikawa, Phys. Rev.
{\bf D 52}, 4985 (1995);
V. Agrawal, T.K. Gaisser, P. Lipari, and T. Stanev, Phys. Rev.
{\bf D 53}, 1314 (1996);
T.K. Gaisser {\it et al.}, Phys. Rev. {\bf D 54}, 5578 (1996);
T.K. Gaisser and T. Stanev, Phys. Rev. {\bf D 57}, 1977 (1998).
%
\bibitem{oldatmos}
J.G. Learned, S. Pakvasa, and T.J. Weiler,
Phys. Lett. {\bf B 207}, 79 (1988);
V. Barger and K. Whisnant, Phys. Lett. {\bf B 209}, 365 (1988);
K. Hidaka, M. Honda, and S. Midorikawa, Phys. Rev. Lett. {\bf 61}, 1537
(1988).
%
\bibitem{barhall}
R. Barbieri, L.J. Hall, D. Smith, A. Strumia, and N. Weiner,
hep-ph/9807235.
%
\bibitem{bw98}
V. Barger and K. Whisnant, hep/ph-9812273, to appear in Phys. Rev. D.
%
\bibitem{bks98}
J.N. Bahcall, P.I. Krastev, and A.Yu. Smirnov,
Phys. Rev. {\bf D 58}, 096016 (1998).
%
\bibitem{bw99}
V. Barger and K. Whisnant, hep-ph/9903262, to appear in Phys. Lett. B.
%
\bibitem{hdm}
See, e.g., J.R. Primack, Science {\bf 280}, 1398 (1998);
J.R. Primack and M.A.K. Gross, astro-ph/9810204;
E. Gawiser and J. Silk, Science {\bf 280}, 1405 (1998).
%
\bibitem{tegmark}
D.J. Eisenstein, W. Hu, and M. Tegmark,
Phys. Rev. Lett. {\bf 80}, 5255 (1998); astro-ph/9807130.
%
\bibitem{lya}
R.A.C. Croft, W. Hu, and R. Dave, astro-ph/9903335.
%
\bibitem{0nubb}
M. Doi {\it et al.}, Phys. Lett. {\bf B 102}, 323 (1981).
%
\bibitem{books}
{\it The Physics of Massive Neutrinos}, B. Kayser, F. Gibrat-Debu, and
F. Perrier (World Scientific, Singapore, 1989);
{\it Massive Neutrinos in Physics and Astrophysics}, R. Mohapatra and
P.B. Pal (World Scientific, Singapore, 1991).
%
\bibitem{majorana}
H. Minakata and O. Yasuda, Phys. Rev. {\bf D 56}, 1692 (1997);
T. Fukuyama, K. Matsuda, and H. Nishiura,
Phys. Rev. {\bf D 57}, 5844 (1998);
H. Georgi and S.L. Glashow, hep-ph/9808293;
G.C. Branco, M.N. Rebelo, and J.I. Silva-Marcos,
Phys. Rev. Lett. {\bf 82}, 683 (1999);
R. Adhikari and G. Rajasekaren, hep-ph/9812361.
%
\bibitem{MSW}
L. Wolfenstein, Phys. Rev. {\bf D 17}, 2369 (1978);
V. Barger, R.J.N. Phillips, and K. Whisnant,
Phys. Rev. {\bf D 22}, 2718 (1980);
P. Langacker, J.P. Leveille, and J. Sheiman,
Phys. Rev. {\bf D 27}, 1228 (1983);
S.P. Mikheyev and A. Smirnov, Yad. Fiz. {\bf 42}, 1441 (1985);
Nuovo Cim. {\bf C 9}, 17 (1986);
S.J. Parke, Phys. Rev. Lett. {\bf 57}, 1275 (1986);
S.J. Parke and T.P. Walker, Phys. Rev. Lett. {\bf 57}, 2322 (1986);
W.C. Haxton, Phys. Rev. Lett. {\bf 57}, 1271 (1986);
T.K. Kuo and J. Pantaleone, Rev. Mod. Phys. {\bf 61}, 937 (1989).
%
\bibitem{vacuum}
V. Barger, R.J.N. Phillips, and K. Whisnant,
Phys. Rev. {\bf D 24}, 538 (1981);
S.L. Glashow and L.M. Krauss, Phys. Lett. {\bf B 190}, 199 (1987);
V. Barger, R.J.N. Phillips, and K. Whisnant,
Phys. Rev. Lett. {\bf 65}, 3084 (1990);
{\it ibid.} {\bf 69}, 3135 (1992);
A. Acker, S. Pakvasa, and J. Pantaleone,
Phys. Rev. {\bf D 43}, 1754 (1991);
P.I. Krastev and S.T. Petcov, Phys. Lett. {\bf B 285}, 85 (1992);
Phys. Rev. {\bf D 53}, 1665 (1996);
N. Hata and P. Langacker, Phys. Rev. {\bf D 56}, 6107 (1997);
S.L. Glashow, P.J. Kernan, and L.M. Krauss,
Phys. Lett. {\bf B 445}, 412 (1999).
%
\bibitem{CHOOZ}
CHOOZ collaboration, M. Apollonio {\it et al.},
Phys. Lett. {\bf B 420}, 397 (1998).
%
\bibitem{Mlimit}
L. Baudis {\it et al.}, hep-ex/9902014.
%
\bibitem{GENIUS}
H.V. Klapdor-Kleingrothaus, J. Hellmig, and M. Hirsch,
J. Phys. {\bf G 24}, 483 (1998).
%
\bibitem{MAP}
See, e.g., http://map.gsfc.nasa.gov
%
\bibitem{PLANCK}
See, e.g., http://astro.estec.esa.nl/SA-general/Projects/Planck
%
\bibitem{SDSS}
See, e.g., http://www.sdss.org
%
\bibitem{2dF}
See, e.g., http://meteor.anu.edu.au/$\sim$colless/2dF
%
\bibitem{tritium}
A.I. Belesev {\it et al.}, Phys. Lett. {\bf B 350}, 263 (1995);
V.M. Lobashev {\it et al.}, in {\it Proc. of Neutrino-96},
ed. by K. Enqvist, K. Huitu, and J. Maalampi (World Scientific,
Singapore, 1997).
%
\bibitem{inferred}
V. Barger, T.J. Weiler, and K. Whisnant,
Phys. Lett. {\bf B 442}, 255 (1998).
%
\bibitem{phases}
$CP$ phases of Majorana neutrinos are discussed in
Ref.~\cite{majorana}; see also
L. Wolfenstein, Phys. Lett. {\bf B 107}, 77 (1981);
P.B. Pal and L. Wolfenstein, Phys. Rev. {\bf D 25}, 766 (1982);
K. Hagiwara and N. Okamura, hep-ph/9811495.
%
\bibitem{K2K}
K2K Collaboration, Y. Oyama, Proc. of the YITP Workshop on Flavor
Physics, Kyoto, Japan, 1998, hep-ex/9803014.
%
\bibitem{MINOS}
MINOS Collaboration, ``Neutrino Oscillation Physics at Fermilab: The
NuMI-MINOS Project,'' NuMI-L-375, May 1998.
%
\end{thebibliography}
\end{document}